\def\year{2022}\relax
\documentclass[letterpaper]{article} 
\usepackage{bm}
\usepackage{amstext}
\usepackage{booktabs}
\usepackage{amsmath}
\usepackage{subfigure}
\usepackage{xcolor}         
\usepackage{aaai22}  
\usepackage{times}  
\usepackage{helvet}  
\usepackage{courier}  
\usepackage[hyphens]{url}  
\usepackage{graphicx} 
\usepackage{natbib}  
\usepackage{caption} 
\DeclareCaptionStyle{ruled}{labelfont=normalfont,labelsep=colon,strut=off} 
\usepackage{algorithm}
\usepackage{algorithmic}

\usepackage{newfloat}
\usepackage{listings}
\floatstyle{ruled}
\newfloat{listing}{tb}{lst}{}
\floatname{listing}{Listing}

\usepackage{array}
\newcolumntype{L}[1]{>{\raggedright\let\newline\\\arraybackslash\hspace{0pt}}m{#1}}
\newcolumntype{C}[1]{>{\centering\let\newline  \\\arraybackslash\hspace{0pt}}m{#1}}
\newcolumntype{R}[1]{>{\raggedleft\let\newline \\\arraybackslash\hspace{0pt}}m{#1}}
\title{Graph Neural Networks for Double-Strand DNA Breaks Prediction}
\author{
    Xu Wang\footnote{Xu Wang is a research intern in 4Paradigm. This paper has been accepted by DLG-AAAI'22.}, \textsuperscript{\rm 1}
    Huan Zhao, \textsuperscript{\rm 1}
    Weiwei Tu, \textsuperscript{\rm 1}
    Hao Li, \textsuperscript{\rm 2}
    Yu Sun, \textsuperscript{\rm 2}
    Xiaochen Bo \textsuperscript{\rm 2}

}
\affiliations{

    \textsuperscript{\rm 1} 4Paradigm\\
    \textsuperscript{\rm 2} Beijing Institute of Radiation Medicine\\
    \{wangxu01,zhaohuan,tuweiwei\}@4paradigm.com,\{lihao,sytower1994,boxiaoc\}@163.com

}

\usepackage{bibentry}

\begin{document}

\maketitle

\begin{abstract}
    Double-strand DNA breaks (DSBs) are a form of DNA damage that can cause abnormal chromosomal rearrangements. Recent technologies based on high-throughput experiments have obvious high costs and technical challenges.
    Therefore, we design a graph neural network based method to predict DSBs (GraphDSB), using DNA sequence features and chromosome structure information. 
    In order to improve the expression ability of the model, we introduce Jumping Knowledge architecture and several effective structural encoding methods.
    The contribution of structural information to the prediction of DSBs is verified by the experiments on datasets from normal human epidermal keratinocytes (NHEK) and chronic myeloid leukemia cell line (K562), and the ablation studies further demonstrate the effectiveness of the designed components in the proposed GraphDSB framework. 
    Finally, we use GNNExplainer to analyze the contribution of node features and topology to DSBs prediction, and proved the high contribution of 5-mer DNA sequence features and two chromatin interaction modes.
\end{abstract}

\section{Introduction}

\begin{figure*}[t]
	\centering
	\includegraphics[width=0.85\textwidth]{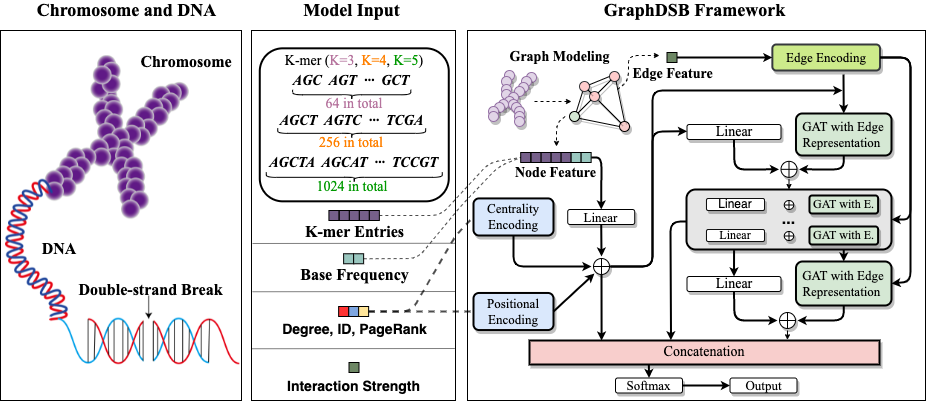}
	\caption{The overall architecture of the proposed GraphDSB, including the illustration of double-strand DNA break and \emph{K-mer} entries, graph modeling of chromosomes, the composition of node features, and the network architecture. The $\bigoplus$ represents the addition operation between matrices. 
}	\label{Figure1}
\vspace{-10pt}
\end{figure*}

Double-strand DNA breaks (DSBs) refer to the situation where both DNA strands of the double helix structures are broken, as shown in the left part of Figure~\ref{Figure1}. 
DSBs are usually caused by the attack of reactive oxygen species and other electrophilic molecules on deoxyribose and DNA bases \cite{mckinnon2007dna}. 
Because the processing and repair of DSBs can lead to mutations, loss of heterozygosity, and chromosomal rearrangement, which can lead to cell death or cancer \cite{mehta2014sources}.
At present, several high-throughput sequencing technologies generally rely on some potential nucleases or sequencing based methods to map human endogenous DSBs with high resolution \cite{lensing2016dsbcapture}.

However, due to high sequencing costs and experimental difficulties, DSBs have only been localized with high resolution in a few cell lines, which has prevented the comprehensive study of the DSB landscape in the human genome across diverse cell lines and tissues \cite{mourad2018predicting}.
At the same time, machine learning plays an increasingly important role in accelerating genomic studies, including integrating gene expression data \cite{chereda2019utilizing, rhee2017hybrid}, characterizing non-coding interactions \cite{zhang2019predicting, wu2020inferring} and classifying diseases across the genome \cite{zhao2020deeplgp}, etc.

Since chromosome structure has been proven to contribute to DSBs prediction \cite{mourad2018predicting}, considering the ability of graph neural network (GNN) \cite{battaglia2018relational} to capture structural information, in this work, we propose a novel GNN framework to predict DSBs (denoted as GraphDSB).
GraphDSB combines the strength of interaction in the calculation of attention coefficients, and introduces Jumping Knowledge (JK) network \cite{xu2018representation}, \emph{centrality encoding} \cite{ying2021transformers} and \emph{positional encoding} \cite{vaswani2017attention} to further improve the expression ability of the model, whose effectiveness are verified in our experiments on two real-world datasets, i.e., the normal cell line NHEK and the cancer cell line K562.
Furthermore, to analysis the results of GraphDSB, we use GNNExplainer \cite{ying2019gnnexplainer} to analyze the contribution of DNA sequence characteristics and chromosome topology to the prediction of DSBs, which proves the importance of 5-mer DNA sequence features and two chromatin interaction modes for DSBs prediction. 

\section{The Proposed Framework}

In this section, we introduce the proposed GraphDSB in detail, which includes graph modeling of chromosomes, a GNN framework based on self-attention \cite{velivckovic2017graph} with edge representation and JK-Network \cite{xu2018representation}, the \emph{centrality encoding} which includes degree and PageRank \cite{page1999pagerank}, and the \emph{positional encoding} which uses node IDs. The overall framework is shown in the right part of Figure \ref{Figure1}. 

\subsection{Graph Modeling of Chromosomes}

 First, in order to use the information provided by chromosome structure and DNA shape, we model the chromosome as a graph structure $G=(V,E)$, in which node $v_{i} \in V$ represents a 5-kb genome bin. The label of each node is represented by $y_{i}\in\{0,1\}$, where $1$ means a double-strand break has occurred, and $0$ means normal. The edges $E$ reflect the interaction strength between these bins, i.e., the contact frequency of corresponding entry in Hi-C contact map we collected, which reflects the chromosome structural information.
 The intention of our model is to fit the mapping relationship between chromosome structure, DNA sequencing results and DSBs, so we define DSBs prediction as a node-level binary classification problem. 
 
In order to save the computing resources, we extract \emph{K-mer} entries as shown in Figure~\ref{Figure1}, where \emph{K-mer} represents the DNA sequence with length \emph{K}, and \emph{K-mer} entries is the frequency of these sequences, which are widely used in the analysis of genome sequence \cite{chor2009genomic}, and calculate their frequency distribution to replace the one-hot coding of AGCT sequence \cite{kurtz2008new}. Furthermore, we introduce base frequency, which refers to the frequency of each bases, i.e.,  $A, G, C, T$,  in the whole genome bin to enrich node features. 
 	\vspace{-5pt}

 \subsection{The GraphDSB Framework}

After modeling the chromosome into a graph, we then design a GNN architecture to aggregate structural information of the graph. 
Let the feature vector of node $v_{i}$ be $\bm{x}_{i}$, for the representation of nodes, modern GNN framework is generally expressed as two steps: aggregation of neighbor messages and update of representation.
Specifically, denote $\bm{h}_{i}^{(l)}$ as the representaiton of $v_{i}$ in the $l$-th layer and define $\bm{h}_{i}^{(0)}=\bm{x}_{i}$. The computation of neighborhood aggregation and update in the $l$-th layer is in the following:
\begin{equation}
    \begin{aligned}
        \bm{m}_{i}^{(l)}=\text{AGG}^{(l)}\left(\left \{ \bm{h}_{j}^{(l-1)}:j\in N(v_{i}) \right \}\right), \\
        \bm{h}_{i}^{(l)}=\text{UPDATE}^{(l)}\left(\bm{h}_{i}^{(l-1)},\bm{m}_{i}^{(l)}\right)\text{,}
    \end{aligned}
    \end{equation}
where $\bm{m}_{i}^{(l)}$ represents the message aggregated from the neighbors of $v_{i}$ in the $l$-th layer, and $N(v_{i})$ is the set of first-order neighbors of $v_{i}$.

\textbf{Graph attention with edge representation.} Tbe original GAT \cite{velivckovic2017graph} does not introduce edge features into the calculation of \emph{attention coefficient}, which is not sufficient to predict DSBs, since edge features contain
important structural information of chromosomes. Therefore, we introduce edge features as follows:
{\scriptsize
\begin{equation}
  \bm{\alpha}_{ij} = \frac{\text{exp}(\text{LeakyReLU}\big(\bm{a}^{T}[\bm{W}_{N}\bm{h}_{i}||\bm{W}_{N}\bm{h}_{j}||\bm{W}_{E}\bm{h}_{(i,j)}])\big)}{\sum _{k\in N(v_{i})}\text{exp}\big(\text{LeakyReLU}(\bm{a}^{T}[\bm{W}_{N}\bm{h}_{i}||\bm{W}_{N}\bm{h}_{k}||\bm{W}_{E}\bm{h}_{(i,k)}])\big)},
\end{equation}
}
where $\bm{a}$ is a learnable weight vector, $\bm{W}_{N}$ and $\bm{W}_{E}$ denote two trainable weight matrix, respectively, and $||$ dentoes the concatenation operation.
The edge feature $\bm{h}_{(i,j)}$ is obtained from the interaction strength $\bm{e}_{(i,j)}$ between two genome bins through a linear layer, which is defined as \emph{edge encoding}. The purpose of \emph{edge encoding} is to keep the edge feature and the node feature dimension consistent. 

To further improve the expression ability of the model, we introduce the JK-Network architecture here, as shown in Figure~\ref{Figure1}. Because the final representation of nodes can be adaptively integrated the information of different layers, we can build a deeper network, so as to achieve better results in DSBs prediction, which is verified in our experiments.

\textbf{Centrality encoding and positional encoding.} Following Graphormer \cite{ying2021transformers}, we define a \emph{centrality encoding}, which not only considers the degree of nodes, but also introduces PageRank. 
At the same time, considering the sequence of chromosome structure, we also introduce the \emph{positional encoding} in Transformer \cite{vaswani2017attention} as a supplement to node features.
Instead of the absolute position of words, we use the ID of the genome bin in the chromosome.
Finally, we multiply them by different learnable matrices $\bm{W}_{deg}$, $\bm{W}_{pr}$ and $\bm{W}_{pos}$, to achieve the same dimension as the node features, and initialize the node representation in a summation way:
$\bm{h}_{i}^{(0)}=\bm{x}_{i}+\bm{z}_{deg(v_{i})}+\bm{z}_{pr(v_{i})}+\bm{z}_{pos(v_{i})}$.
By adding the \emph{centrality encoding} and the \emph{positional encoding} to the input, the model can capture both the semantic correlation and the node importance in the attention mechanism.

Since the task is to identify DSBs on chromosomes (node-level binary classification), we use cross-entropy as loss function.

\begin{table}[h]
	\footnotesize
	\caption{Statistics of the two datasets in the experiments.}
	\label{dataset}
	\centering
	\begin{tabular}{cC{30pt}C{36pt}C{50pt}C{33pt}}
		\toprule
		 Dataset  & Avg. \#Nodes &Avg. \#Edges & Density(\%) & Avg. P/N ratio \\
		\midrule
		NHEK & 24,123 & 1,126,839 & 0.489 & 6.26    \\
		K562 & 24,079 & 3,743,946 & 1.784 & 107.35     \\
		\bottomrule
	\end{tabular}
\vspace{-10pt}
\end{table}

\section{Experiments}

\textbf{Settings.} 
We utilize the chromosomes in two human cells, respectively normal human epidermal keratinocytes (NHEK) and chronic myeloid leukemia cell line (K562) datasets\footnote{The DSB datasets for NHEK and K562 are available at the NCBI (https://www.ncbi.nlm.nih.gov/) under accession code GSE78172 and Sequence Read Archive at SRP099132.}. Among them, NHEK is a normal cell line, which is available from the epidermis, it is the major cell type in the epidermis (making up about 90\%). K562 is a cancer cell line. Specifically, K562 cells were established as the first human immortalized myelogenous leukemia line. These two cell lines are often used as representatives for erythroleukemic normal and tumour cell respectively. Each of dataset contains 22 autosomes and X chromosome (we do not consider Y chromosome).

For each cell line, we gathered its (1) DSB dataset as well as (2) the Hi-C dataset which reflects the chromosome structural information. The DNA sequence is derived from the human reference genome.
Because DSBs are more common in cancer cells, there is also a large gap in the positive-negative ratio (P/N ratio), as shown in Table \ref{dataset}.

For each dataset, we use 21 chromosomes as training, 1 as validation, and the remaining one as test, which is presented in the form of inductive learning. For the evaluation metric, we use Area Under Curve (AUC).

All models are implemented with Pytorch (version 1.7.0) \cite{paszke2019pytorch} on a GPU 2080Ti. In addition, in order to facilitate the implementation of various GNN variants, we use the popular GNN library: Deep Graph Library (DGL) (version 0.6.0) \cite{wang2019deep}.

\begin{table}[h]
		\footnotesize
	\caption{Comparisons among different models in terms of average AUC with STDs. The best performance is in bold.}
	\label{performance}
	\centering
	\begin{tabular}{ccc}
		\toprule
		Method     & NHEK & K562 \\
		\midrule
		LightGBM & $0.7507\pm0.0299$    & $0.7397\pm0.0525$  \\
		MLP      & $0.7623\pm0.0291$    & $0.7237\pm0.0528$  \\
		GraphDSB  & $\bm{0.7902\pm0.0233}$   & \bm{$0.7564\pm0.0443$}  \\
		\bottomrule
	\end{tabular}
\end{table}

\textbf{Performance.} In order to evaluate the contribution of the graph structure to the identification of DSBs, we select two effective models, LightGBM \cite{ke2017lightgbm} and Multilyaer Perceptron (MLP), which do not depend on the graph structure, as the baselines.
Our GNN model includes 1 input layer $(d=256)$, 4 GNN layers $(d=256)$, 1 full connection layer $(d=256\times5)$ and 1 output layer$(d=1)$.
The number of attention heads is set to 4. The MLP architecture is also five-layered, and the dimensions are consistent with GraphDSB. 
For LightGBM, we set reasonable parameter search field for each hyperparameter, and use Hyperopt \cite{bergstra2013hyperopt} for automatic search, the number of serach iterations is setted to 100. For all NN method, We use Adam as the optimizer, the learning rate is set to 0.003, the exponential decay strategy is adopted 
and the rate is 0.99. 
The results are shown in Table \ref{performance}. 

In terms of average AUC, GraphDSB performs better than the other two methods that do not leverage on graph structure, which demonstrates the effectiveness of the graph modeling, i.e., chromosomes structures, on the prediction of DSBs.

\begin{table}[t]
		\footnotesize
	\caption{Ablation studies on the proposed GraphDSB.}
	\label{ablation}
	\centering
	\begin{tabular}{ccc}
		\toprule
		Method         & NHEK & K562 \\
		\midrule
		w/o JK                          & $0.7799\pm0.0027$  & $0.7451\pm0.0044$    \\
		w/o E.R.         & $0.7831\pm0.0026$  & $0.7472\pm0.0045$    \\
		w/o C.E.  & $0.7886\pm0.0026$  & $0.7428\pm0.0045$    \\
        w/o P.E.  & $0.7893\pm0.0028$  & $0.7531\pm0.0047$    \\
		\midrule
		\midrule
		GCN            & $0.7699\pm0.0027$   & $0.7353\pm0.0046$  \\
		GAT            & $0.7805\pm0.0027$   & $0.7448\pm0.0044$  \\
		GIN            & $0.7361\pm0.0029$   & $0.7361\pm0.0046$  \\
		\midrule
		\midrule
		GraphDSB            & \bm{$0.7902\pm0.0023$}  & \bm{$0.7564\pm0.0044$}   \\
		\bottomrule
	\end{tabular}
  \vspace{-0.4cm}
\end{table}

\textbf{Ablation study.} 
The ablation study results are given in Table \ref{ablation}, including whether JK is included, whether interaction strength are introduced, whether \emph{centrality encoding} and \emph{positional encoding} is introduced, and the comparison with popular GNN variants, including GCN \cite{kipf2016semi}, GAT \cite{velivckovic2017graph} and GIN \cite{xu2018powerful}.
It can be seen that GAT performs better than the other two GNN methods, which proves the role of the attention mechanism from the side.
While GraphDSB is better than these three GNN methods, it is also better than the four variants of GraphDSB, which verifies our original intention of designing the network framework, \emph{centrality encoding} and \emph{positional encoding}. 

\textbf{Explaining the results of GraphDSB.}
In order to analyze the contribution of node features and topologies to DSBs prediction, we introduce GNNExplainer, its core idea is to use the mutual information $MI(Y,(G_{s},X_{s}))$ between input substructures $G_{s}$ and sub features $X_{s}$ to measure the importance of node features and edges.

\begin{figure}[h]
    \centering
    \subfigure[Node features importance]{
      \label{node_features}
      \includegraphics[scale=0.25]{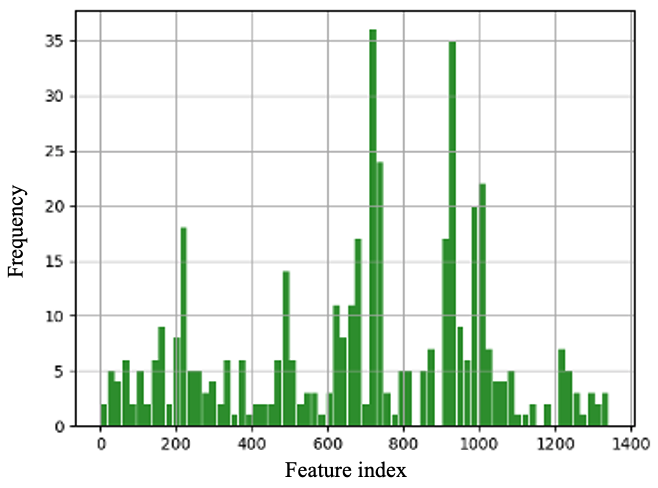}
    }
\centering
    \subfigure[The most frequent motifs]{
      \label{motif1+2}
      \includegraphics[scale=0.25]{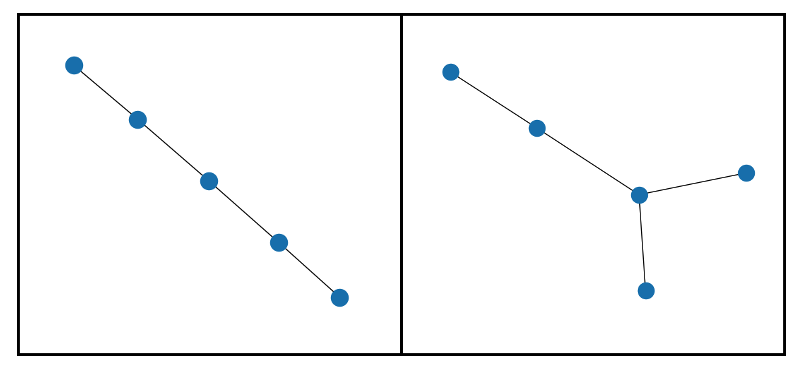}
    }
    \caption{The results of GNNExplainer application on GraphDSB. (a) The distribution of the importance of the node feature. The TOP20 feature ID is counted and recorded on each chromosome. The vertical axis indicates the frequency of the feature ID in the TOP20. In order to facilitate the display, the 1344-dimensional features are grouped into 70 bins. (b) The most frequent motifs, the left is the 'forward chain' and the right is the 'binary-parallel'.}
    \label{explain}
    \vspace{-10pt}
  \end{figure}

For all chromosomes, we first calculate the feature importance of all nodes and take the average value, and final count the TOP 20 feature numbers with the largest average value of importance. As shown in Figure \ref{node_features}, we found that 5-mer DNA sequence features(the index is between 321-1344) are more important than 3-mer or 4-mer DNA sequence features, suggesting the complex DNA sequence preference of DSBs.

Beyond important genomic features, we extract the importance of edges which provides a set of crucial chromatin interactions that are most influential to predict whether DSB occurs at a given genome site of interest. These interactions form a functional module that corresponds to a series of genome regions that have pair-wise interactions nearby the prediction site. 
We referred to this module as DSB associated chromatin interaction module (DaCIM) for that prediction site.
We limit the edge importance calculation to 2-hop neighbor, and found a group of motifs: recurring and significant patterns of chromatin interactions in DaCIM. 
The most frequent motifs tend to have a mode termed 'forward chain' and 'binary-parallel' Figure \ref{motif1+2}. These motifs suggests DaCIM may  contain the universal build block at interaction level for chromatin organization.
Similar discoveries have also been made in \cite{falk2010higher}, which discussed the role of higher-order chromatin structure in DSB induction and repair.


\section{Related Work}

DSB is generally detected based on high-throughput experiments. For example, \citep{crosetto2013nucleotide} proposed a method of direct in situ break labeling, streptavidin enrichment and next-generation sequencing to draw DSBs with nucleotide resolution. There are other methods, such as \cite{biernacka2018bless,keimling2008sensitive}.
Although high-throughput technology allows genome-wide mapping of DSBs with high resolution, high cost and high technical difficulty are still unavoidable problems. 
Recently, more and more studies have focused on low-cost, high-efficiency DSBs prediciton.
\cite{mourad2018predicting} demonstrate, that endogenous DSBs can be computationally predicted using the epigenomic and chromatin context,   
and it used Random Forest as a model for predicting DSBs. 
However, compared to the proposed GraphDSB, this method does not use the structural information of chromosomes when predicting DSBs, and some studies have shown that the shape of DNA has a certain correlation with DSBs \cite{taverna2007chromatin, zhou2015predicting, mathelier2016dna}.

Recent years have witnessed the effectiveness of graph neural network on graph-structured data. Representative GNNs, including graph convolutional network (GCN) \cite{kipf2016semi}, GraphSAGE \cite{hamilton2017inductive}, graph attention network (GAT) \cite{velivckovic2017graph}, 
graph isomorphsim network (GIN) \cite{xu2018powerful}, have been widely used for different graph-based tasks.
For GNN, graph structured inputs offer representational flexibility, which can naturally model proteins, moleclues and so on. Therefore, GNN has also made significant development in biology and genomics in recent years \cite{ingraham2019generative, fout2017protein, nguyen2021graphdta, wang2021property}. However, to the best of our knowledge, in this work, we make the first attempt to explore the ability of GNN on the prediction of DSBs by modeling the chromosome structure.

\section{Conclusion}

In this work, we explore the GraphDSB to DSBs prediction, which uses DNA sequence features and chromosome structure information as input.
GraphDSB performes better than other GNN variants on the two datasets of the normal cell line NHEK and the cancer cell line K562, which verifies the effectiveness of the proposed framework and several structural encodings methods.
We also used GNNExplainer to prove the effectiveness of 5-mer DNA sequence features and found two important interaction modes for DSBs prediction. In the future, we will explore the neural architecture search method and apply the existing work to the field of biomedicine\cite{zhang2021automated, huan2021search, wei2021pooling, wei2021designing}.


\bibliography{reference}

\end{document}